\documentclass[aps,12pt,preprintnumbers]{revtex4}
\usepackage{amsmath,amssymb,bm,epsfig}
\renewcommand\d{\partial}

\renewcommand\r{\mathbf{r}}
\newcommand\p{\mathbf{p}}
\newcommand\<{\langle}
\renewcommand\>{\rangle}

\begin{document}
\preprint{INT-PUB 07-12}
\title{Three comments on the Fermi gas at unitarity in a harmonic trap}
\author{D.~T.~Son}
\affiliation{Institute for Nuclear Theory, University of Washington,
Seattle, Washington 98195, USA}
\date{June 2007}
\begin{abstract}
\noindent
In this note we consider three issues related to the unitary Fermi gas
in a harmonic trap.  We present a short proof of a virial theorem,
which states that the average energy of a particle system at unitarity
in a harmonic trap is twice larger than the average potential energy.
The theorem is valid for all systems with no intrinsic scale, at zero
or finite temperature.  We discuss the odd-even splitting in a
unitarity Fermi gas in a harmonic trap.  We show that at large number
of particles $N$ the odd-even splitting is proportional to
$N^{1/9}\hbar\omega$, with an undetermined numerical constant.  We
also show that for odd $N$ the lowest excitation energies are of order
$N^{-1/3}\hbar\omega$.

\end{abstract}
\maketitle

Recently Fermi gas at unitarity has been realized and studied
experimentally.  In most experimental setups the system is confined in
a harmonic trap, hence of great interest are the properties of such a
system in such traps.  In this note we explore several aspects of this
system: the virial theorem, the odd-even splitting, and the excited
energy levels of a system with an odd number of particles.

\bigskip

{\bf 1.}  Reference~\cite{Thomas} contains a proof of a virial theorem for a
unitary Fermi gas in a harmomic trap.  This theorem states that the
total energy is exactly twice the average potential energy due to the
trap,
$\< H\> = 2\< V\>$.
The proof is based on the force balance in a harmonic
trap within the local density approximation.
There is, however, a simpler proof based on the Hellmann-Feynman theorem.
It is similar to, but arguably simpler, than the a proof due to Chevy quoted 
in Ref.~\cite{WernerCastin}.


Consider the spherical trap first.  The Hamiltonian of the system can
be written as
\begin{equation}\label{H}
  H(\omega) = \sum_{a=1}^N
  \left( \frac{\p_{a}^2}{2m} + \frac{m\omega^2 \r_{a}^2}2\right)
  + H_{\textrm{2-body}}.
\end{equation}
where $H_{\textrm{2-body}}$ includes all two-body interactions.  We
have highlighted the fact that the Hamiltonian depends on the trap
frequency $\omega$.

Since in the unitarity limit the system lacks an intrinsic scale, the
ground state energy has to be proportional to $\hbar\omega$, which is
the only energy scale available.  Denote the ground state of $N$
particles as $|\omega,N\>$, we have
\begin{equation}\label{scaling}
  \<\omega,N |H(\omega)|\omega,N\> = c_N \hbar\omega,
\end{equation}
where $c_N$ is a constant dependent on $N$.

The Hellmann-Feynman theorem states that
\begin{equation}\label{Feynman}
  \omega\frac{\d}{\d\omega} \<\omega,N |H(\omega)|\omega,N\>
  = \<\omega,N |\omega\frac{\d H(\omega)}{\d\omega}
    |\omega,N
    \>.
\end{equation}
We see from Eq.~(\ref{scaling}) that the left-hand side is equal to
the ground state energy, and from Eq.~(\ref{H}) that the right-hand
size is equal to twice the average potential energy.  One obtains
the virial theorem: $\<H\>=2\<V\>$.

It is easy to extend the theorem to the case of an anisotropic trap.
In this case the Hamiltonian is
\begin{equation}
  H = \sum_{a=1}^N\sum_{i=1}^3
  \left( \frac{\p_{ai}^2}{2m} + \frac{m\omega_i^2\r_{ai}^2}2\right)
  + H_{\rm pair},
\end{equation}
and instead of Eq.~(\ref{scaling}) we have
\begin{equation}
  \<\omega_i, N|H(\omega_i)|\omega_i, N\> = E_N(\omega_i).
\end{equation}
Consider a generalization of Eq.~(\ref{Feynman}):
\begin{equation}\label{HF-anharm}
   \sum_i \omega_i\frac{\d}{\d\omega_i}\<\omega_i, N| H(\omega_i)
   |\omega_i, N\> =
   \<\omega_i, N|\sum_i\omega_i\frac{\d H(\omega_i)}{\d\omega_i} 
   |\omega_i, N\>.
\end{equation}
Basic dimensionality analysis tells us that $E_N(\omega_i)$ is equal
to a common frequency, e.g., the geometric mean
$(\omega_1\omega_2\omega_3)^{1/3}$, times a dimensionless function of
the ratios between frequencies.  Such a function is homogeneous
function of first order of its arguments, i.e.,
\begin{equation}
  \sum_{i=1}^3 \omega_i \frac{\d}{\d\omega_i}E_N(\omega_i) = E_N(\omega_i).
\end{equation}
while the Hamiltonian itself is a homogeneous function of second order
with respect to $\omega_i$.  From Eq.~(\ref{HF-anharm}) we obtain the
virial theorem in an anharmonic trap.

The proof can be extended to finite temperature $T$ as well.  For
simplicity consider the harmonic trap.  The free energy has to scale as
\begin{equation}\label{scaling-T}
  F_N(\omega, T) = c_N\left(\frac T\omega\right) \hbar\omega .
\end{equation}
Now $c_N$ can depend on the dimensionless ratio $T/\omega$.  We now
apply the finite-temperature version of the Hellmann-Feynman formula,
\begin{equation}
  \omega \frac{\d}{\d\omega}F_N(\omega, T) = 
  \< \omega\frac{\d H}{\d\omega}\>.
\end{equation}
The right hand size is again $2\<V\>$.  The left hand side, from
Eq.~(\ref{scaling-T}), can be transformed to
\begin{equation}
  \omega \frac{\d F_N}{\d\omega} = F_N - T\frac{\d F_N}{\d T}
  = F + TS,
\end{equation}
which, according to a thermodynamic relation, the average energy
$\<H\>$ in the canonical ensemble.  The extension to anharmonic traps
and grand canonical ensemble is straightforward.

We note that the proof does not rely on the local density
approximation.  Compared to Chevy's proof (quoted in
Ref.~\cite{WernerCastin}), here we rescale the potential while in
Chevy's proof the wavefunction is rescaled.  The theorem holds for
systems with spin imbalanced, independent of their phase structure.

Finally we note that the virial theorem should be valid for other
systems with no intrinsic scales, for example for the system of anyons
in two-dimensional harmonic traps.

\bigskip

{\bf 2.} The quantum Monte-Carlo calculation of
Ref.~\cite{BertschChang} shows a clear even-odd effect, reminiscent of
the behavior of the nuclear binding energy.  If one denotes by $E_N$
the ground energy of $N$ particles in an isotropic harmonic trap
($E_N=c_N\hbar\omega$) then the quantity
\begin{equation}
  \Delta_N = E_{N} - \frac12 (E_{N-1} + E_{N+1}), 
\end{equation}
for odd $N$, is positive and approximately $0.6-0.7\hbar\omega$ for
the range of particle number $N$ studied (3 to 21).  Recall that the
ground state for even $N$ has equal numbers of spin-up and spin-down
fermions, while for odd $N$ the numbers differ by one.

The effect is clearly related to pairing between particles, but
a question arises: will the odd-even effect remains constant when one
goes to large $N$?  Here we argue that in the limit of large $N$, the
even-odd splitting has to grow as a small power of $N$:
\begin{equation}\label{DeltaN-adv}
  \Delta_N = E_{N} - \frac12 (E_{N-1} + E_{N+1}) \sim N^{1/9}\omega, \qquad
  \textrm{$N$ large and odd}
\end{equation}

To see how does this dependence arise, first recall that in the local
density approximation (LDA) the system, with an even number of
particles, can be thought of as a Fermi gas with a spatially dependent
chemical potential:
\begin{equation}
  \mu(\r) = \mu_0 - V(\r) 
  = \mu_0 - \frac12 m\omega^2 \r^2
\end{equation}
(we consider an isotropic trap).  Assume we have a system with even
$N$, and discuss the process of introducing an extra particle into it.
For $N\gg1$, the extra particle can be thought of as an extra
fermionic quasiparticle carrying definite spin.
If $N\gg1$ the cloud is large, and 
the quasiparticle can be approximately localized at a position $\r$.  The
energy cost of doing so is the chemical potential, plus the local
energy gap $\Delta(\r)$ (due to superfluidity), which is proportional
to $\mu$:
\begin{equation}
  \Delta(\r) = C\mu(\r)
\end{equation}
($C\approx1.2$ according to a Monte-Carlo
simulation~\cite{Carlson:2005kg}).  Therefore the extra particle will
be localized where the gap is smallest, which is near the edge of the
cloud, $r=R$.  Unfortunately, the LDA breaks down there.

Since the quasiparticle is localized on the thin shell, the 
curvature of the shell can be completely ignored.  As the result, the
problem is mapped onto the problem of inserting a quasiparticle in a
symmetric system at $\mu=0$ in the linear potential
\begin{equation}\label{Ez}
  V(\r) = {\cal E}z 
\end{equation}
where 
\begin{equation}\label{Efield}
  {\cal E}=m\omega^2R.
\end{equation}
The choice of the notation reflects the
fact that ${\cal E}$ can be interpreted as an electric field.

\begin{figure}[ht]
\includegraphics[width=0.4\textwidth]{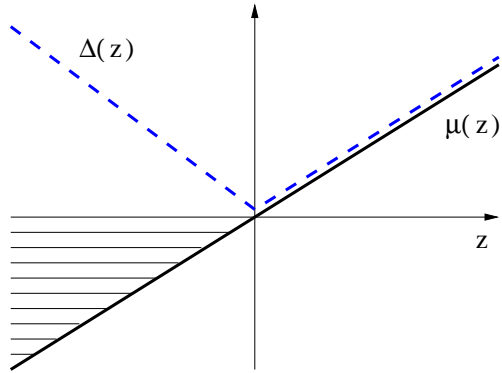}
\caption[]{Localization of an extra unpaired quasiparticle on the
surface of the cloud.  The solid line is the effective chemical
potential, and the dashed line is the effective potential experienced
by an extra unpaired quasiparticle.  The picture is not
quantitatively accurate near the edge of the cloud.}
\label{fig:localization}
\end{figure}

In the potential~(\ref{Ez}) the symmetric system, in the local density
approximation, fills the $z<0$ half of space; the other half $z>0$ is
empty.  Beyond LDA there is a boundary layer around $z=0$.  The
density tends to infinity as $z\to-\infty$ but it is fine for a
scale-free system as the unitary Fermi gas.  Assuming ${\cal E}>0$,
the quasiparticle cannot stray toward negative $z$, because the gap
there is large, and it cannot stray toward positive $z$ because of the
potential.  Therefore it has to be localized at the boundary $z=0$
(Fig.~\ref{fig:localization}).  At unitarity the ground state energy
of this quasiparticle can be estimated by dimensional analysis by
constructing the unique combination of $\hbar$, $m$, and ${\cal E}$
with the dimension of energy,
\begin{equation}\label{Eextra}
  E_{\rm extra} = \chi \frac{(\hbar{\cal E})^{2/3}}{m^{1/3}}\,.
\end{equation}
Here $\chi$ is a universal dimensionless constant.

Recall that the size of the cloud $R$ is related to the particle
number $N$ by
\begin{equation}\label{RN}
  R = {\xi^{1/4}} \sqrt{\frac\hbar{m\omega}}\, (24N)^{1/6}
\end{equation}
where $\xi$ is an universal number (defined as the ratio of energy of
an unitary Fermi gas and the energy of a free Fermi gas with the same
density), we find, by combining Eqs.~(\ref{Efield}), (\ref{Eextra}),
and (\ref{RN}),
\begin{equation}
  \Delta_N = {\chi}{\xi^{1/6}}(24N)^{1/9}\hbar\omega
\end{equation}
which is the behavior advertised in Eq.~(\ref{DeltaN-adv}).  Thus we
conclude that the odd-even splitting grows as a small power of $N$ at
large $N$.

While $\xi$ has been evaluated by many methods (which typically give
$\xi\approx0.4$~\cite{Arnold:2006fr}), we have no previous evaluation
of $\chi$.  Ideally, one would like to use quantum Monte-Carlo method
to find it, but it should be also possible to estimate $\chi$ by using
the $\epsilon$ expansion technique~\cite{Nishida:2006br}.  One could
use the Bogoliubov-de Gennes equation to find a rough estimate of
$\chi$.  This is, however, beyond the scope of this note.

We can also discuss the odd-even splitting in an anharmonic trap.  We take
the limit of large $N$, fixed aspect ratios of the trap.  The
extra particle is still located near the edge of the cloud, but
instead of spreading out over the whole ellipsoidal edge it will be
attracted to the areas where the potential gradient (the effective
electric field ${\cal E}$) is smallest.  This corresponds the
locations farthest from the center. Repeating the calculations we shall
find
\begin{equation}
  \Delta_N = \chi\xi^{1/6}(24N)^{1/9}\hbar\bar\omega^{1/3}
  \omega_{\rm min}^{2/3}
\end{equation}
where $\bar\omega=(\omega_1\omega_2\omega_3)^{1/3}$ and $\omega_{\rm
min}$ is the minimal among $\omega_1$, $\omega_2$, and $\omega_3$.

\bigskip

{\bf 3.} Let us now discuss the excited energy levels of a system in a
harmonic trap.  For a system with even $N$ and equal numbers of
spin-up and spin-down fermions, the lowest-energy excitation can be
thought of as a coherent excitation of the superfluid, and the lowest
energy excitations have energies of order
$\hbar\omega$~\cite{BaranovPetrov}.

For an odd system, another type of excitations exist.  Recall that we
now have a extra quasiparticle moving on the surface of the cloud.  
We assume that this quasiparticle has the dispersion relation 
$\epsilon=k^2/2m^*$ at small momentum $k$ along the surface, where $m^*$ is
an effective mass.  In
the ground state the extra particle has momentum zero.  One can give
this quasiparticle a nonzero orbital momentum, as the result one get a
series of excitation level whose energy depends on the orbital
momentum $\ell$ as
\begin{equation}
  E_\ell = \frac{\hbar^2\ell(\ell+1)}{2m^*R^2} = 
  \frac{\ell(\ell+1)}{2\xi^{1/2}}\frac1{(24N)^{1/3}}
  \frac m{m^*} \hbar\omega
\end{equation}
What is interesting is that the splitting between lowest energy levels
is not given by the scale $\hbar\omega$ as in even systems, but by a
much smaller scale $N^{-1/3}\hbar\omega$.

It is also interesting to discuss the excitations of an odd system in
an anharmonic trap.  Again we consider the limit of large $N$ at fixed
trap aspect ratios.  If the smallest frequency is unique (e.g.., if
$\omega_1<\omega_2$ and $\omega_1<\omega_3$) then the extra particle
is attracted to two opposite points on the longest axis of the
ellipsoidal cloud.  When the cloud is large the particle can jump
between the two points only by quantum tunneling.  Thus for odd number
of particles $N$ the ground state is almost degenerate with the first
excited state, which has an opposite parity.  This ``parity doubling''
does not hold when the trap is an oblate spheroid (e.g., if
$\omega_1=\omega_2<\omega_3$), since now the extra particle is
concentrated to a circular ring around the cloud edge.

I am indebted to G.~Bertsch, A.~Bulgac, S.~Y.~Chang, J.~E.~Drut,
M.~Forbes, A.~Kryjevski,
and S.~Tan for discussions leading to this note, and to
J.~Thomas for comments on the manuscript.  This work is supported, in
part, by DOE Grant No.\ DE-FG02-00ER41132.


\begin{thebibliography}{9}


\bibitem{Thomas}
  J.~E.~Thomas, J.~Kinast, and A.~Turlapov,
  ``Virial theorem and universality of a unitary Fermi gas,''
  Phys.\ Rev.\ Lett.\ {\bf 95}, 120402 (2005) [arXiv:cond-mat/0503620].

\bibitem{WernerCastin}
  F.~Werner and Y.~Castin,
  ``Unitary gas in an isotropic harmonic trap: symmetry properties and
  applications,''
  Phys. Rev. A {\bf 74}, 053604 (2006) [cond-mat/0607821].

\bibitem{BertschChang}
  S.~Y.~Chang and G.~F.~Bertsch,
  ``Unitarity Fermi gas in a harmonic trap,''
  physics/0703190.


\bibitem{Arnold:2006fr}
  See, e.g., a summary in 
  P.~Arnold, J.~E.~Drut, and D.~T.~Son,
  ``Next-to-next-to-leading order $\epsilon$ expansion for a Fermi gas 
  at infinite scattering length,''
  Phys. Rev. A {\bf 75}, 043505 (2007)
  [arXiv:cond-mat/0608477.

\bibitem{Carlson:2005kg}
  J.~Carlson and S.~Reddy,
  ``Asymmetric two-component fermion systems in strong coupling,''
  Phys.\ Rev.\ Lett.\  {\bf 95}, 060401 (2005)
  [arXiv:cond-mat/0503256].

\bibitem{Nishida:2006br}
  Y.~Nishida and D.~T.~Son,
  ``$\epsilon$ expansion for a Fermi gas at infinite scattering length,''
  Phys.\ Rev.\ Lett.\  {\bf 97}, 050403 (2006)
  [arXiv:cond-mat/0604500].

\bibitem{BaranovPetrov}
  M.~A.~Baranov and D.~S.~Petrov,
  ``Low-energy collective excitations in a superfluid trapped Fermi gas,''
  Phys.\ Rev.\ A {\bf 62}, 041601 (2000)
  [arXiv:cond-mat/9901108].

\end{thebibliography}
\end{document}